\documentclass[letterpaper, 10pt, conference]{ieeeconf}  

\IEEEoverridecommandlockouts                              

\usepackage{graphics} 
\usepackage{epsfig} 
\usepackage{mathptmx} 
\usepackage{times} 
\usepackage{amsmath} 
\usepackage{amssymb}  
\usepackage[dvipsnames]{xcolor}
\usepackage{soul}

\title{\LARGE \bf
Brain-Computer Interfaces: Investigating the Transition from Visually Evoked to Purely Imagined Steady-State Potentials}

\author{Arturo Micheli$^{1}$, Davide Consoli$^{2}$, Adrien Merlini$^{3}$, Paolo Ricci$^{4}$, and Francesco P. Andriulli$^{5}$
\thanks{*This work was supported in part by the European Research Council (ERC) through the European Union’s Horizon 2020 Research and Innovation Programme under Grant 724846 (Project 321) and in part by the Italian Ministry of University and Research within the Program FARE, CELER, under Grant R187PMFXA4.}
\thanks{$^{1}$Arturo Micheli is a former research assistant at the Department of Electronics and Telecommunications, Politecnico di Torino, 10129 Turin, Italy
        {\tt\small arturo.micheli@polito.it}}%
\thanks{$^{2}$Davide Consoli is a PhD student at the Department of Electronics and Telecommunications, Politecnico di Torino, 10129 Turin, Italy
        {\tt\small davide.consoli@polito.it}}%
\thanks{$^{3}$Adrien Merlini is an Associate Professor at the Microwave Department, IMT Atlantique, 29238 Brest, France
        {\tt\small adrien.merlini@imt-atlantique.fr}}%
\thanks{$^{4}$Paolo Ricci is a PhD student at the Department of Electronics and Telecommunications, Politecnico di Torino, 10129 Turin, Italy
        {\tt\small paolo.ricci@polito.it}}%
\thanks{$^{5}$Francesco Andriulli is a Full Professor at the Department of Electronics and Telecommunications, Politecnico di Torino, 10129 Turin, Italy
        {\tt\small francesco.andriulli@polito.it}}%
}

\begin{document}

\maketitle
\thispagestyle{empty}
\pagestyle{empty}

\begin{abstract}
Brain-Computer Interfaces (BCIs) based on Steady State Visually Evoked Potentials (SSVEPs) have proven effective and provide significant accuracy and information-transfer rates. This family of strategies, however, requires external devices that provide the frequency stimuli required by the technique. This limits the scenarios in which they can be applied, especially when compared to other BCI approaches. In this work, we have investigated the possibility of obtaining frequency responses in the EEG output based on the pure visual imagination of SSVEP-eliciting stimuli. Our results show that not only that EEG signals present frequency-specific peaks related to the frequency the user is focusing on, but also that promising classification accuracy can be achieved, paving the way for a robust and reliable visual imagery BCI modality.
\newline

\indent \textit{Clinical relevance}---Brain computer interfaces play a fundamental role in enhancing the quality of life of patients with severe motor impairments. Strategies based on purely imagined stimuli, like the one presented here, are particularly impacting, especially in the most severe cases.
\end{abstract}


\section{INTRODUCTION}
Brain-computer interfaces (BCIs) aim at obtaining non-muscular channels to send information to external devices \cite{nicolas}, and their clinical relevance spans several fields including the improvement of the quality of life for patients with motor impairments of diverse degrees of severity \cite{lazarou}. Among non-invasive BCI approaches, substantial research has been devoted to systems based on electroencephalography (EEG). As in other strategies, EEG-based BCIs generally follow a standard workflow: signal acquisition, pre-processing and signal enhancement, feature extraction and classification, and translation of the features into commands.

Among common EEG-based BCI approaches, several studies have focused on investigating steady-state visually evoked potentials (SSVEPs) \cite{guger,wu,zhu}. In this paradigm, a visual stimulus, often a pattern flickering at a specific frequency \cite{wolpaw}, results in peaks in the power spectral density (PSD) of the EEG acquisitions, both at the frequency of the stimulus and at its higher order harmonics. When a set of different panels flickering at different frequencies are presented to the user---who selects and concentrates only on one of them---a PSD analysis can discriminate among the panels and find the selected one, which results in a usable transfer of information with significant accuracy and high bit-rates \cite{chen}. 
This notwithstanding, standard SSVEP techniques come with some drawbacks. Since SSVEP entirely relies on an external stimulus, subjects need to continuously gaze at a screen to be able to communicate. This may not always be possible in all application scenarios, for example when the users suffer \textcolor{black}{from ocular motor dysfunction} or very severe forms of motor impairments.

In this work, to overcome the limitations of standard SSVEP-based BCIs, we propose and analyze a new visual imagery (VI) paradigm, based on imagining SSVEP-eliciting stimuli and investigating the feasibility of BCIs based on the analysis of the corresponding EEG signals. Preliminary tests confirm the the new paradigm is suitable for real-time applications including moving a cursor on a screen and selecting elements from a grid. In Section~\ref{sec:methods} we briefly describe our experimental setting, the methods and the strategies adopted, while Section~\ref{sec:discussion} presents our results and discussion. Finally, Section~\ref{sec:conclusions} delineates our conclusions and venues for further investigations.

\section{METHODS}\label{sec:methods}

Visual imagery can be defined as the representation of information related to perception in the absence of retinal inputs \cite{kaski}. On the contrary, visual perception (VP) is derived from the acquisition of visual information through the eyes \cite{kosslyn}. The objective of this work is to propose a new form of VI, investigating whether mental representations of flickering patterns could provide a frequency-specific power increase (as examined through PSD analysis), similar to those obtained for SSVEPs.
\textcolor{black}{Several studies on SSVEP-based BCIs and related datasets have been presented in the literature \cite{acampora2021dataset,wang2016benchmark}, but to the best of the authors' knowledge, they do not include studies on VI-based BCIs relying on the proposed paradigm.}

\begin{figure}
\centering
\includegraphics[width=0.9\linewidth]{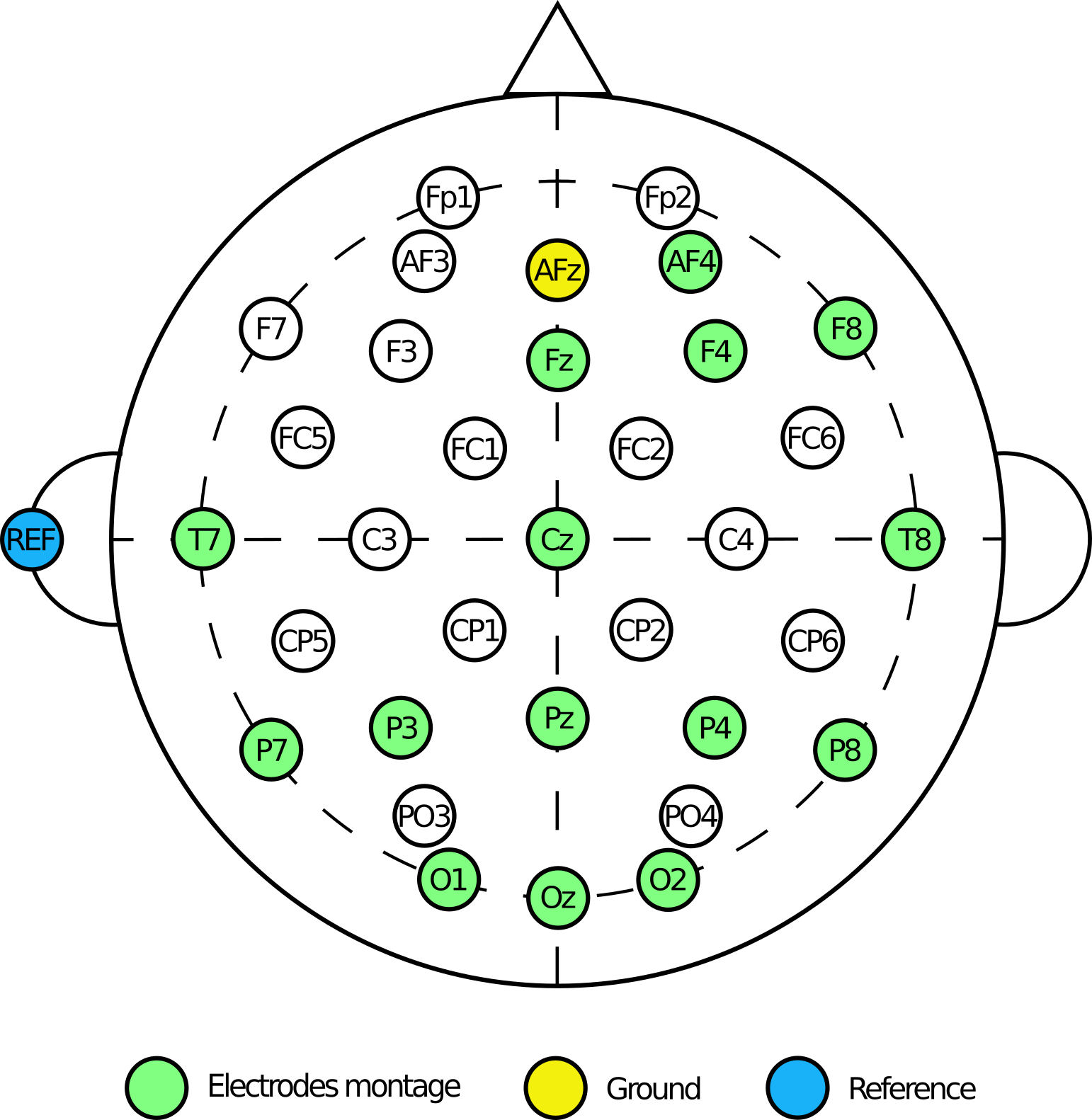}
\caption[Electrodes positioning]{Electrode positioning on the EEG cap, using the international $10-20$ system. Our signal acquisition electrodes are indicated in green, while yellow and blue indicate the the ground and reference electrodes, respectively.}
\label{fig:electrodepos}
\end{figure}

The experimental setting consisted of a 16 \textcolor{black}{active wet} electrodes EEG device, \textcolor{black}{working with a sampling frequency of 256Hz,} a monitor showing chessboard flickering patterns with controllable frequency, and of two speakers producing auditory stimuli of controllable switch frequency. The origin of the VI signal is still a matter of discussion and various regions of activation have been suggested. The most relevant ones are the occipital areas (associated with the similarity of VI with VP) \cite{sabbah} and the parietal and frontal networks \cite{kosmyna}, with a particular attention to the right-frontal zone \cite{azmy}. This dictated our choice of principal locations for our EEG acquisitions. Following the literature we have selected: AF4, F4, F8, O1, O2, Pz, P3, P4, Fz, Cz, Oz, T7, T8, P7 and P8 (refer to Fig.~\ref{fig:electrodepos} for electrodes positioning configuration). 

\textcolor{black}{To train a subject to drive a BCI via VI signals, w}e followed three protocols. In the first protocol, experiments were lead using one single frequency at a time, either $5$~Hz or $7$~Hz. These frequencies were chosen because they were sufficiently low to be easily imagined but still distant enough from each other to be easily classified. In the protocol, SSVEP trials were classified together with VI trials and rest. In VI trials, the subject was asked to imagine SSVEP-eliciting visual stimuli. This protocol was conceived as a concurrent VI-SSVEP one, precisely to frequently refresh the user's memory of a SSVEP stimulus via the SSVEP trials. In some experiments of this first protocol ($1.a$ and $1.b$ below) we provided an additional support to the user: an auditory co-stimulus during VI trials, which was beeping at the same frequency at which VI had to be conducted. These acoustic stimuli were then removed once the subjects were feeling more confident ($1.c$ and $1.d$). Experiments without acoustic stimuli always followed experiments with acoustic stimuli. 

In the second protocol, experiments were lead using two frequencies at a time, $5$~Hz and $7$~Hz. As in the previous protocol, SSVEP trials were classified together with VI trials and rest. Like before in the VI trials, the subject was asked to imagine SSVEP-eliciting visual stimuli by imagining SSVEP trials. In this protocol no auditory co-stimuli were provided to the subject. This second protocol always followed the sessions of the previous one.

In the third and final protocol, experiments were lead using two frequencies at a time, $5$~Hz and $7$~Hz. Differently from previous protocols, however, no SSVEP trials were classified together with VI trials and rest and no auditory co-stimuli were provided. This third protocol always followed the sessions of the two previous ones.

For the sake of clarity, the three protocols are summarized below. The first 3 numbers represent the trial duration (in seconds), the amount of trials per each class, and the number of classes for the specific session, respectively.
\begin{enumerate}
    \item \textbf{Single frequency concurrent VI and SSVEP} $-$ Concurrent classification of SSVEP, VI, and rest at one single frequency 
    \begin{enumerate}
        \item $6s\times15\times3$ at $5$~Hz, with SSVEP, VI, rest, with auditory co-stimuli during VI trials;
        \item $6s\times15\times3$ at $7$~Hz, with SSVEP, VI, rest, with auditory co-stimuli during VI trials;
        \item $6s\times15\times3$ at $5$~Hz, with SSVEP, VI, rest;
        \item $6s\times15\times3$ at $7$~Hz, with SSVEP, VI, rest;
    \end{enumerate}
    \item \textbf{Multiple frequencies concurrent VI and SSVEP} $-$ Concurrent classification of SSVEP, VI and rest at two frequencies 
    \begin{enumerate}
        \item $6s\times18\times5$, with $5$~Hz on the left side of the monitor and $7$~Hz on the right side (for SSVEPs only), including SSVEP, VI, and rest;
    \end{enumerate}
    \item \textbf{Multiple frequencies VI only} $-$ Concurrent classification of VI and rest classes at two frequencies 
    \begin{enumerate}
        \item $9s\times20\times3$, with $5$~Hz and $7$~Hz, including only VI and rest trials;   
    \end{enumerate}
\end{enumerate}
From the above summary it is clear that trials lasted from 6 to 9 seconds, depending on the test conducted. In sessions with visual stimuli, 6 seconds were considered sufficient to elicit a response. For pure VI experiments instead, we opted to increase the duration,  to facilitate the observation of the studied signal. Trials were proposed to the subject in a random order, so as to prevent the the user from anticipating the task he needed to perform. Two-frequencies experiments were implemented with an arrow pointing to the frequency the user had to concentrate on.

Signals obtained from the acquisition sessions were processed through a $60$~Hz low-pass filter for electromyographic signal and high frequency noise attenuation, and a $48$--$52$~Hz notch filter, for the rejection of the line noise. Then, a $8^\text{th}$ order Butterworth filter was implemented, with bandpass frequencies in the range $2$--$36$~Hz. From the raw data collected, only windows of 4 seconds were utilized for classification, specifically from second 2 to second 6 for the first two protocols and from second 3 to second 7 for the third protocol. PSDs estimations were performed on the windowed signals for every electrode. After this processing we collected for each trial a data vector containing the average power spectral density for each electrode in the range $2$--$36$~Hz. These vectors were fed as features to the classifier. 

For all the experiments, classification was performed via a regularized support vector machine (SVM) classifier with linear kernel. The whole classification pipeline presents two hyper-parameters. The first one is the subset of electrodes to be adopted for the classification. The second one is the regularization parameter of the SVM classifier. For this reason, before the actual training of the model, we implemented a heuristic algorithm that tunes the two hyper-parameters through the splitting in two of the training data-set. Once the hyper-parameters were optimized, we used the whole training data-set to train our model, and this was utilized to classify the testing data-set. Multiple recording sessions have been used for the realization of the two data-sets, so as to grant significant dimensions for both training and testing sets. Obviously, none of the sessions were in common between the two types of data-sets.

\section{RESULTS AND DISCUSSION}\label{sec:discussion}

\begin{figure}
\centering
\includegraphics[width=\linewidth]{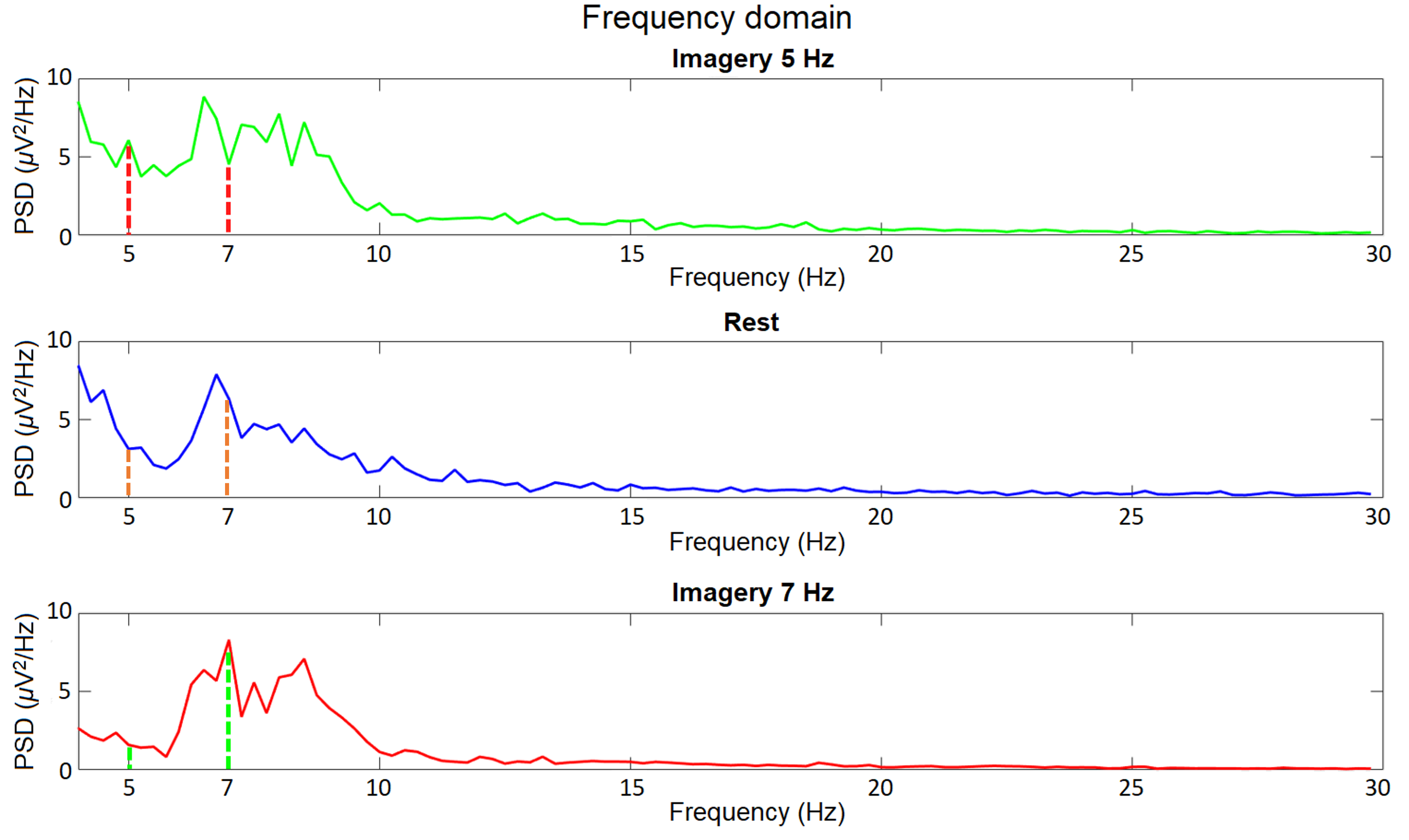}
\caption[Averaged PSD of VI trials for multiple frequencies experiments]{Averaged PSD of the signals regarding the sessions involving visual imagination and rest only, acquired from the AF4 electrode. Peaks can be seen at $5$~Hz and at $7$~Hz.}
\label{fig:only_vi}
\end{figure}

Performing the power spectral density analysis of the processed recordings and averaging among trials, we could gather visual results of the various sessions tested, from single frequency to multiple frequencies experiments. Our preliminary experiments were conducted on one subject from our research group (male, 23 years old).
Our results are reported in Fig.~\ref{fig:only_vi}, where the frequency-specific peaks of activation for a purely imagined session with multiple frequencies involved ($5$~Hz and $7$~Hz) are evidenced. As Fig.~\ref{fig:only_vi} is showing, not only did the trials related to a certain frequency show an increase in the PSD amplitude, but also a substantial decrease of the other corresponding rate (e.g. the $7$~Hz one for $5$~Hz trials and vice-versa) was evident. 

\begin{figure}
\centering
\includegraphics[width=\linewidth]{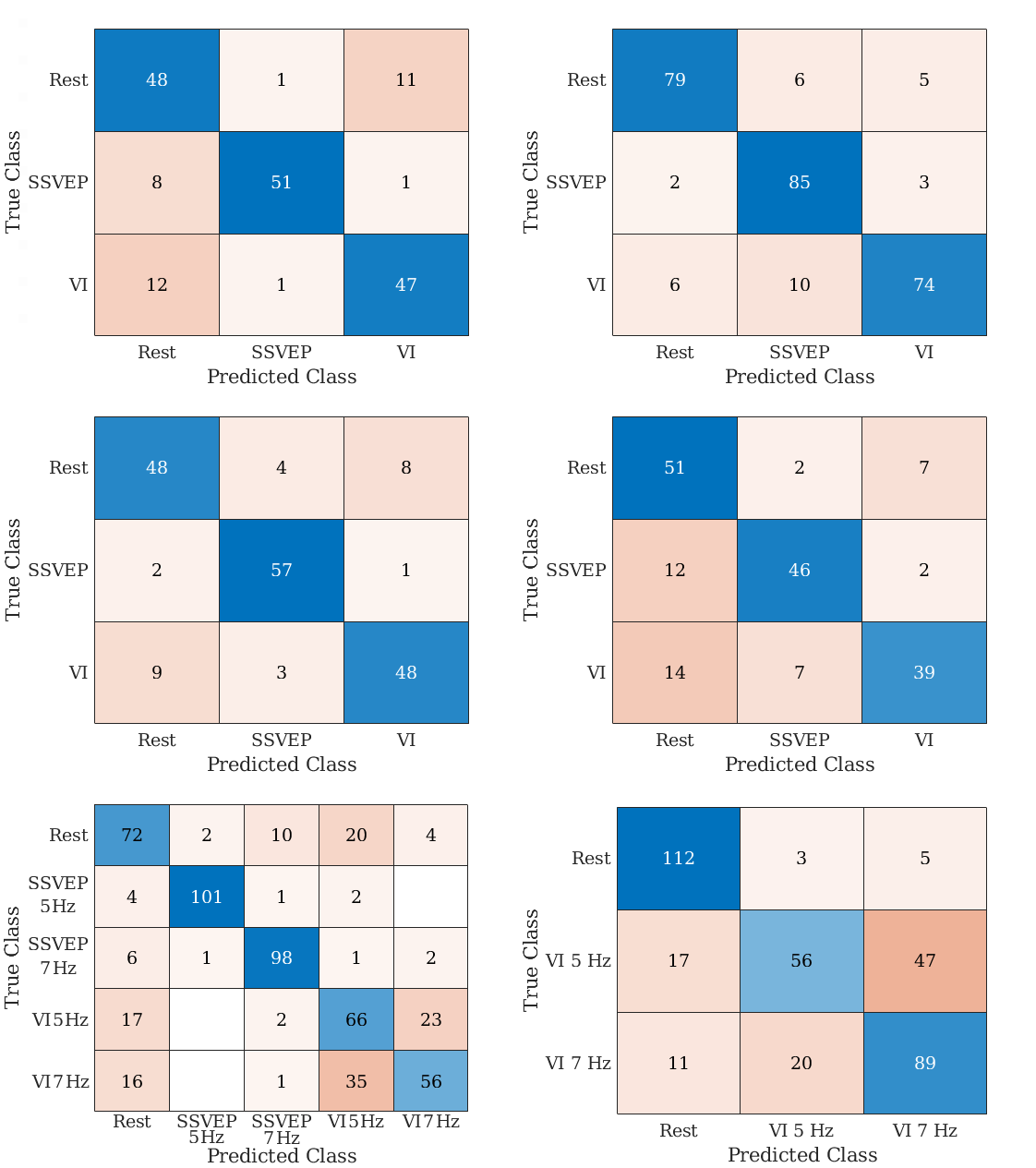}
\caption{Confusion matrices of the classification results shown in Table \ref{tableI}. Experiments ordered from top left to bottom right follow the order used for the Table.}
\label{fig:confMat}
\end{figure}

Then, the signals have been classified via SVM. Table~\ref{tableI} reports the accuracy for the various classification sessions, compared with standard SSVEP. Initial sessions at single frequency, concerning protocol phases $1.a-1.b$ (see Section~\ref{sec:methods}), lead to success rate of above $81\%$ on testing. When the auditory co-stimuli were removed, in protocol phases $1.c-1.d$ (see Section~\ref{sec:methods}), a classification accuracy of at least $75\%$ could be obtained in each experiment. In single frequency experiments, some of the sessions under-performed. This is most probably due to the fact that in these tests we can observe that the model is over-fitting, leaving margins for future improvements. In two-frequencies sessions (protocol phase $2.a$, see Section~\ref{sec:methods}), the accuracy attained for the test set was of $72.78\%$. In the experiments including only visually imagined patterns and rest (protocol phase $3.a$, see Section~\ref{sec:methods}), an accuracy of $71.39\%$ was obtained. The last two columns of Table~\ref{tableI} report the total number of trials adopted for classification purposes on the different sessions. As is clearly noticeable, even in the most difficult condition (pure VI), a percentage above $70\%$ could be reached. This value is significantly higher than the condition related to random classification ($33.3\%$ for 3 classes experiments), and this is promising for the use of the VI signals proposed here for BCIs. \textcolor{black}{In this setting the offline bit-rate is approximately 4 bits/min, following the definition of \cite{bitrate}.} For visualization purposes, Figure~\ref{fig:confMat} reports the confusion matrices of the results obtained for the different sessions.

Moreover, to validate the robustness of our paradigm, tests were conducted on the same subject after a period of time of 9 months during which the user did not perform further experiments. The new recordings were used as testing data-set, while as training data-set we used all previously recorded data, obtaining results that are shown in Table~\ref{tableII}. As is clearly visible from the Table, not only did the percentages remain constant throughout the months, but even increased in most of the cases. This was especially evidenced in single frequency sessions without auditory co-stimuli. Increases of $6.11\%$ and of $11.11\%$ could be seen for $5$~Hz and $7$~Hz sessions respectively. This suggests that a BCI based on VI signals would not require an extensive and continuous training. 
 
The results presented here are very preliminary and have to be confirmed by following the protocol on a wider cohort of subjects; this extended study is currently underway.

\begin{table}
\caption{Classification accuracy on training and testing sets, ``Train/Test acc.'' and relative size of data-sets in number of trials, ``Train/Test \#''. Table legend: ``$5$-$7$~Hz w/ s.'': single frequency with auditory co-stimuli; ``$5$-$7$~Hz w/o s.'': single frequency without auditory co-stimuli; ``Mult. freq.'': multiple frequencies with SSVEPs; ``Pure VI'': only VI multiple frequencies and rest, no visual stimuli.}
\label{tableI}
\begin{center}
\begin{tabular}{|c||c||c||c||c|}
\hline
\textbf{Session}  & \textbf{Train acc.}  & \textbf{Test acc.}  & \textbf{Train \#} & \textbf{Test \#} \\ \hline
\textbf{$5$~Hz w/ s.} & $100.00\%$ & $81.11$\% & 450 & 180 \\ \hline
\textbf{$7$~Hz w/ s.} & $96.48\%$ & $88.15$\% & 540 & 270 \\ \hline
\textbf{$5$~Hz w/o s.} & $94.72\%$ & $85.00$\% & 360 & 180 \\ \hline
\textbf{$7$~Hz w/o s.} & $100.00\%$ & $75.56$\% & 450 & 180 \\ \hline
\textbf{Mult. freq.} & $91.67\%$ & $72.78$\% & 1080 & 540 \\  \hline
\textbf{Pure VI} & $75.77\%$ & $71.39$\% & 780 & 360 \\ \hline
\end{tabular}
\end{center}
\end{table}

\begin{table}
\caption{Average classification accuracy for tests on the same subject after 9 months without training. For Table~\ref{tableII} legend please refer to Table~\ref{tableI}.}
\label{tableII}
\begin{center}
\begin{tabular}{|c||c||c||c||c|}
\hline
\textbf{Session}  & \textbf{Train acc.}  & \textbf{Test acc.}  & \textbf{Train \#} & \textbf{Test \#} \\ \hline
\textbf{$5$~Hz w/ s.} & $100.00\%$ & $82.22$\% & 630 & 45 \\ \hline
\textbf{$7$~Hz w/ s.} & $95.80\%$ & $88.89$\% & 810 & 45 \\ \hline
\textbf{$5$~Hz w/o s.} & $93.15\%$ & $91.11$\% & 540 & 45 \\ \hline
\textbf{$7$~Hz w/o s.} & $100.00\%$ & $86.67$\% & 630 & 45 \\ \hline
\textbf{Mult. freq.} & $89.38\%$ & $72.22$\% & 1620 & 90 \\  \hline
\textbf{Pure VI} & $76.93\%$ & $73.33$\% & 1140 & 60 \\ \hline
\end{tabular}
\end{center}
\end{table}

\section{CONCLUSIONS}\label{sec:conclusions}

In conclusion, in this work experiments aimed at a preliminary evaluation of the feasibility of a BCI based purely on VI signals have been carried out. For each of our different sessions, encouraging results were gathered from our BCI pipeline, both in terms of visualization of the signals and of classification accuracy. Thus, the testing of our approach on a wider cohort of subjects together with further investigations in the development of a BCI driven by purely imagined flickering patterns are warranted.
Future developments will also include \textcolor{black}{tests in which several frequencies are used within the same experiment (thus extending the paradigm's number of commands and bit-rate),} the implementation of an online session for more realistic BCI studies, statistical tests on more subjects, and optimization of our spatial filter using inverse source approaches, comparing the classification accuracy before and after processing. 

\addtolength{\textheight}{-13.5cm}   



\section*{APPENDIX}
\noindent ALS = Amyotrophic Lateral Sclerosis\\
BCI = Brain-Computer Interface\\
EEG = Electroencephalography \\
PSD = Power Spectral Density\\
SSVEP = Steady-State Visually Evoked Potential\\
SVM = Support Vector Machine\\
VI = Visual Imagery\\
VP = Visual Perception\\


\end{document}